\documentclass{article}

\usepackage[firstpage]{draftwatermark}
\usepackage[T1]{fontenc} 
\usepackage[utf8]{inputenc} 
\usepackage{amsmath,cite,url}
\usepackage{graphicx}
\usepackage{color}
\usepackage[lbd]{ismir}

\definecolor{lightgray}{rgb}{0.9,0.9,0.9}
\definecolor{darkgray}{rgb}{0.4,0.4,0.4}
\SetWatermarkFontSize{12pt}
\SetWatermarkScale{1.1}
\SetWatermarkAngle{90}
\SetWatermarkHorCenter{202mm}
\SetWatermarkVerCenter{170mm}
\SetWatermarkColor{darkgray}
\SetWatermarkText{Late-Breaking / Demo Session Extended Abstract, ISMIR 2025 Conference}

\title{Osu2MIR: Beat Tracking Dataset Derived From Osu! Data}

\twoauthors
  {Ziyun Liu} {Carnegie Mellon University}
  {Chris Donahue} {Carnegie Mellon University}

\def\authorname{Z. Liu, C. Donahue}

\usepackage[bookmarks=false,hidelinks]{hyperref}

\begin{document}

\maketitle

\begin{abstract}

In this work, we explore the use of Osu!, a community-based rhythm game, as an alternative source of beat and downbeat annotations. Osu! beatmaps are created and refined by a large, diverse community and span underrepresented genres such as anime, Vocaloid, and video game music. We introduce a pipeline for extracting annotations from Osu! beatmaps and partition them into meaningful subsets. Through manual analysis, we find that beatmaps with a single timing point or widely spaced multiple timing points ($\geq$5 seconds apart) provide reliable annotations, while closely spaced timing points ($<$5 seconds apart) often require additional curation. We also observe high consistency across multiple annotations of the same song. This study demonstrates the potential of Osu! data as a scalable, diverse, and community-driven resource for MIR research. We release our pipeline and a high-quality subset \textit{osu2beat2025} to support further exploration: \url{https://github.com/ziyunliu4444/osu2mir}.

\end{abstract}
\section{Introduction}\label{sec:introduction}

In recent years, deep learning-based beat tracking models have become dominant, yet their progress is limited by the quality and scope of the available datasets. Traditionally, beats and downbeats are labeled by experts, resulting in datasets that are small in size, genre-limited, and slow to scale with new releases. These datasets are typically annotated by a single individual, which introduces potential bias, and often contain undetected errors that go unnoticed without careful review\cite{beatthis}\cite{BeatNet+}.

Rhythm games are a genre of video games where players perform actions in sync with music. Recently, there has been an increasing number of projects using rhythm game data, but most of them focus on playable chart generation instead of MIR\cite{DDC}\cite{TaikoNation}\cite{osumania}. So far, ADTOF\cite{ADTOF} is the only well-known MIR dataset based on rhythm game charts and is used in drum transcription.

To address these limitations, we explore Osu!, a community-based rhythm game where players create and share beatmaps synchronized to music. Osu! currently hosts over 40,000 beatmap sets contributed by a global community, featuring underrepresented genres such as anime, Vocaloid, and video game music. This crowd-sourced content offers a promising alternative to traditional datasets, benefiting from continuous updates. Furthermore, multiple annotations for a single song help reduce individual bias, and community consensus mechanisms contribute to improved annotation quality.

In this work, we introduce ways to construct high-quality beat and downbeat datasets from Osu! beatmaps, outline data collection and processing pipelines, and evaluate the reliability and characteristics of the resulting annotations. Finally, we release a high-quality subset with 741 annotations for 708 distinct audios for direct access.




\section{Data Collection}

Batch Beatmap Downloader \cite{downloader} is a tool developed by an Osu! player nzbasic that allows users to download large numbers of osu! beatmap sets efficiently and is used in our data collection process.

In this study, we applied a series of filters within the downloader to ensure the quality and rhythmic complexity of the selected beatmaps. We restricted our selection to ranked beatmap sets in Osu! mode, the most popular mode in the game. Ranked beatmaps were chosen because they meet the ranking criteria and have gone through the ranking process, ensuring high data quality. To ensure rhythmic complexity, we included only beatmap sets that contain "Insane" difficulty levels, which represent the second-highest difficulty tier in the game. Lastly, we restricted our dataset to beatmap sets with at least 200 favorites.

\section{Data Processing}

Information related to timing is stored in \texttt{.osu} files, which are plain text files found within each osu! beatmap folder. These files are divided into several sections, one of which is \texttt{[TimingPoints]}, which hosts both uninherited and inherited timing points.

To extract structured beat annotations from the Osu! beatmap files (\texttt{.osz}), we wrote a Python script that automates the extraction and processing of relevant musical metadata and timing information. Each \texttt{.osz} archive was decompressed to retrieve \texttt{.osu} metadata files and associated audio files (\texttt{.mp3}). To ensure correct alignment, the script parses the \texttt{[General]} section of the \texttt{.osu} file to identify the specific audio file referenced by the beatmap. It then parses the \texttt{[TimingPoints]} section to identify uninherited timing points, which give tempo and meter information.

For each uninherited timing point, the script calculates beat timestamps by iteratively adding the beat length until the next uninherited timing point or the end of the song. Beat indices (i.e., positions within each measure) are also computed using the meter. The resulting beat times and indices are stored in a tab-separated annotation file. In parallel, the corresponding audio file is copied to a target directory for further analysis.

\section{Data Partitioning}

In this study, we partitioned the collected data into three subsets based on the structure of uninherited timing points within each beatmap: (1) beatmaps containing only a single uninherited timing point, (2) beatmaps with multiple uninherited timing points spaced at least 5 seconds apart, and (3) beatmaps with multiple uninherited timing points spaced less than 5 seconds apart. The 5-second threshold was chosen as a heuristic to distinguish between distinct timing sections and minor tempo fluctuations within a single section.

\section{Data Analysis}

For each subset, we performed a series of steps to assess the quality of the data.

First, we applied the RNN beat and downbeat tracker in the madmom\cite{madmom} library to the audio files corresponding to each beatmap. The model’s predictions were then compared against ground truth labels derived from the uninherited timing points. We used standard evaluation metrics, including F-measure, CMLt, and AMLt, to quantify the alignment between predicted and annotated beats. These metrics provide a useful but imperfect proxy of data quality.


To more thoroughly assess the quality of the beat and downbeat annotations derived from Osu! beatmaps, we then used madmom evaluation as a heuristic to flag potentially unusual patterns for manual analysis. We manually reviewed a diverse set of examples where madmom’s beat and downbeat evaluations showed low or inconsistent agreement with user annotations. To evaluate the consistency of user annotations, we also selected audio tracks that had been annotated by multiple users and analyzed the degree of agreement between the annotations. Finally, we performed targeted checks based on the specific characteristics we observed in different subsets.

\section{Results}

\textbf{Single timing point}: Songs with one uninherited timing point are rhythmically simple, having mean scores of 0.91 (beat f1), 0.82 (beat cmlt), 0.92 (beat amlt), 0.87 (downbeat f1), 0.80 (downbeat cmlt), and 0.91 (downbeat amlt) when compared to madmom. Manual checks confirm high quality even for syncopated or non-4/4 examples. Multiple annotators show strong agreement. These beatmaps are mostly reliable, but may lack the rhythmic diversity to improve models.

\textbf{Multiple timing points ($\geq$5s apart)}: This subset includes complex rhythms and deliberate meter changes, having mean scores of 0.89 (beat f1), 0.77 (beat cmlt), 0.87 (beat amlt), 0.81 (downbeat f1), 0.72 (downbeat cmlt), and 0.82 (downbeat amlt) when compared to madmom. Annotations align well with the music. Annotators often agree on tempo/meter, though interpretations (e.g., tempo halving) sometimes vary. This group offers high-quality annotations and greater diversity, making it valuable for future research.

\textbf{Multiple timing points ($<$5s apart)}: This subset has mean scores of 0.86 (beat f1), 0.71 (beat cmlt), 0.83 (beat amlt), 0.76 (downbeat f1), 0.61 (downbeat cmlt), and 0.73 (downbeat amlt) when compared to madmom and requires extra filtering and caution. Beatmaps here show high annotation density, often due to expressive timing or wrong meters. Limitations in the Osu! editor (no support for non-4-based meters) and inconsistent annotation goals (e.g., syncing to lyrics) reduce quality.

\section{Pipeline and Dataset Release}

The pipelines provided in our repository can be customized to suit individual research needs. The dataset \textit{osu2beat2025} includes metered beat annotations for the second subset used in our study. The naming format is \texttt{MD5\_BeatmapSetID\_beats\_metered.txt}. Some labels may extend beyond the audible range. Trimming audios to the effective end could be considered when building your own dataset. When annotators use different meters for the same song, the version employing a compound or additive meter is typically the correct interpretation.

\section{Future Directions}

In the data analysis process, we observed that user annotations tend to occur slightly ahead of the beat onset times predicted by madmom, which may reflect how human annotators perceive and internalize rhythm.

In addition, the unique characteristics of Osu! data, such as its inclusion of Japanese, Korean, and Vocaloid songs, present promising opportunities for future research. Differences in language may influence the alignment between lyrics and beats, and Vocaloid tracks offer a distinct opportunity to study artificially generated vocals, which can achieve lyric-music alignments not possible with human singers.

Looking forward, we envision a broader community effort for a bidirectional relationship: MIR enhancing beatmap creation experience, and beatmaps serving as resources for MIR. Plans include developing tools for beatmap creators, developing pipelines for using Osu! data in various MIR tasks, and fostering an open-source ecosystem that enables ongoing refinement and expansion by the community.

\section{Acknowledgments}

We would like to thank Tung-Cheng Su and Purusottam Samal for their contributions to an earlier version of the work.

%

\bibliography{ISMIRtemplate}

%
%
%
%
%

\end{document}